\documentstyle[preprint,eqsecnum,aps]{revtex}
\begin{document}
\tighten
\def\qed{\hfill \vrule height 7pt width 7pt depth 0pt \smallskip}
\newcommand{\ra}{\rightarrow} \newcommand{\al}{\alpha} \newcommand{\sq}{\sqrt}
\newcommand{\hf}{\frac{1}{2}} \newcommand{\iy}{\infty} \newcommand{\la}{\lambda}
\newcommand{\bq}{\begin{equation}} \newcommand{\eq}{\end{equation}}
\newcommand{\qr}{\frac{1}{4}} \renewcommand{\pl}{\partial}
\newcommand{\fr}{\frac}
\def\up#1{\leavevmode \raise .3ex\hbox{$#1$}}
\def\down#1{\leavevmode \lower .5ex\hbox{$\scriptstyle#1$}}
\def\chJ{\up{\chi}_{\down{J}}}
\def\sdown#1{\leavevmode \lower .2ex\hbox{$\scriptstyle#1$}}
\title{Variance Calculations and the Bessel Kernel}
\author{Estelle L.~Basor\footnotemark[1]}
\footnotetext[1]{e-mail address: ebasor@oboe.calpoly.edu}
\address{Department of Mathematics,\\
California Polytechnic State University,
San Luis Obispo, CA 93407, USA}
\author{Craig A.~Tracy\footnotemark[2]}
\footnotetext[2]{e-mail address: tracy@itd.ucdavis.edu}
\address{Department of Mathematics and Institute of Theoretical Dynamics,\\
University of California,
Davis, CA 95616, USA}
\maketitle
\begin{abstract}
In the Laguerre ensemble of $N\times N$ hermitian matrices, it is of
interest
both theoretically and
 for applications to quantum transport problems to compute
the variance of a linear statistic,
denoted ${\rm var}_N~f$,  as $N\ra\iy$.  Furthermore, this
statistic often contains an additional parameter $\al$ for which the
limit $\al\ra\iy$ is most interesting and most difficult to compute
numerically. We derive exact expressions for both 
$\lim_{N\ra\iy} {\rm var}_N~f$ and $\lim_{\al\ra\iy}\lim_{N\ra\iy}
{\rm  var}_N~f$.
\end{abstract}
\section{Introduction and Statement of Main Result}
In the random matrix theory of quantum transport
(see  \cite{beenakker1,beenakker2,stone}
and references therein)  the following
quantity is fundamental:\footnotemark[1]
\begin{eqnarray}
{\rm var}_N f :&=& \int_0^\infty f^2\left(\fr{\sq{4N\mu}}{\al}\right) K_N(\mu,
\mu)\, d\mu \nonumber \\
&&  - \int_0^\infty \int_0^\iy f\left(\fr{\sq{4N\mu}}{\al}\right)
f\left(\fr{\sq{4N\mu'}}{\al}\right) K_N^2(\mu,\mu')\, d\mu d\mu'
\label{varN}
\end{eqnarray}
\footnotetext[1]{The notational choice $f(\sq{x})$ rather than
$f(x)$ in (\ref{varN}) will be convenient later.  It also agrees
with the convention of Stone, et.~al.~\cite{stone}.}
where $K_N(\mu,\mu')$ is the Laguerre kernel; that is
\[
K_N(\mu,\mu')=\sum_{j=0}^{N-1} \phi_j(\mu) \phi_j(\mu')
\]
and $\left\lbrace\phi_j(x)\right\rbrace$ is the sequence of functions
obtained by orthonormalizing the sequence
\[ \left\lbrace x^j x^{\nu/2} e^{-x/2} \right\rbrace_{j=0}^\iy \]
over $(0,\iy)$ (here $\nu>-1$). In particular, one is interested in 
\bq
{\rm var}~f := \lim_{N\ra\iy} {\rm var}_N f
\label{var}
\eq
in the limit  $\al\ra\iy$. In applications various choices
are made for $f$, but we need assume here only that $f$ is smooth
and sufficiently decreasing at infinity to make the integrals well-defined.
\par
In the random matrix model of disordered conductors, the quantity
${\rm var}~f$ is related (via the two-probe Landauer formula) to the
fluctuations of the conductance  and the limit $\al\ra\iy$ is
the high-density (or metallic) regime. A lucid account
can be found in the  
review article by Stone, Mello, Muttalib, and Pichard \cite{stone} to
which we refer the reader for further details and references.
However, these authors did not evaluate ${\rm var}_N f$ in the limits
of interest; namely $N\ra\iy$ followed by $\al\ra\iy$.
It is the purpose of this paper to evaluate  these
limits.  We will see that the result   agrees with the prediction
of Beenakker \cite{beenakker1,beenakker2}
 who gave a heuristic argument for this limit.
\par
By a change of variables
we write (\ref{varN}) in the more suggestive form:
\begin{eqnarray}
{\rm var}_N f &=& \int_0^\iy f^2\left(\fr{\sq{x}}{\al}\right) 
\fr{1}{4N} K_N\left(\fr{x}{4N},\fr{x}{4N}\right) \, dx \nonumber  \\
&& - \int_0^\iy\int_0^\iy f\left(\fr{\sq{x}}{\al}\right)
f\left(\fr{\sq{y}}{\al}\right)
 \left(\fr{1}{4N} K_N\left(\fr{x}{4N},\fr{y}{4N}\right)\right)^2 \, dx dy\, .
\label{varN2}
\end{eqnarray}
From  asymptotic formulas for generalized Laguerre polynomials
(see, e.g., 10.15.2 in \cite{erdelyi})  it follows
that \cite{forrester,tw3}
\begin{eqnarray}
K(x,y):&=&\lim_{N\ra\iy}\frac{1}{4N}K_{N}(\frac{x}{4N},\frac{y}{4N})
\nonumber \\
&=&\,\frac{ J_{\nu}(\sqrt{x})\,\sqrt{y} J_{\nu}'(\sqrt{y})\,-
\sqrt{x} J_{\nu}'(\sqrt{x})\, J_{\nu}(\sqrt{y})}{2(x-y)}
\label{besselK}\end{eqnarray}
where $J_\nu(z)$ is the Bessel function of order $\nu$. (The limit
is uniform in $x$ and $y$ for $0<x,y\leq L < \iy$
and all $L$.) \ \ We call $K(x,y)$ the ``Bessel kernel.'' (This kernel
also arises in scaling the Jacobi ensemble of random matrices at
either edge $\pm 1$.) \ \  
Using this  in (\ref{varN2}) we obtain 
\begin{eqnarray}
{\rm var}~f &=& \int_0^\iy f^2\left(\fr{\sq{x}}{\al}\right) K(x,x)\, dx
\nonumber \\
&& - \int_0^\iy\int_0^\iy f\left(\fr{\sq{x}}{\al}\right)
f\left(\fr{\sq{y}}{\al}\right) K^2(x,y) \, dx dy 
\label{var2}
\end{eqnarray}
where $K$ is the Bessel kernel.
\par
The problem is  reduced to evaluating (\ref{var2}) in the limit
$\al\ra\iy$.  We will show that
\bq
\lim_{\al\ra\iy} {\rm var}~f = \fr{1}{\pi^2}\int_{-\iy}^\iy
\biggl\vert \hat f(2iy)\biggr\vert^2 y \tanh (\pi y) \, dy
\label{constant}
\eq
where $\hat f$ is the Mellin transform of $f$, i.e.
\[ \hat f(z) = \int_0^\iy x^{z-1} f(x)\, dx .\]
This agrees with the result of Beenakker \cite{beenakker1,beenakker2} once one
notes that his $f(x)$ is our $f(\sq{x})$.
For numerous applications of (\ref{constant}) we refer the reader
to Beenakker \cite{beenakker2}. 
\section{The limit $\al\ra\iy$}
\subsection{Use of Hankel Transform}
It is convenient to define the kernel
\begin{eqnarray}
 L(x,y) :&= &2 K(x^2,y^2) \nonumber \\
&=&\int_0^1 t J_\nu(x t) J_\nu(y t)\, dt
\label{L(x,y)}
\end{eqnarray}
where $K(x,y)$ is the Bessel kernel.  (A simple proof of the second
equality can be found in \cite{tw3}.)\ \  Then ${\rm var}~f$ can be
written
\bq {\rm var}~f = \int_0^\iy x f^2\left(\fr{x}{\al}\right) L(x,x)\, dx - I_1 
\label{var3}\eq
where
\begin{eqnarray}
I_1&=&\int_0^\iy \int_0^\iy x y f\left(\fr{x}{\al}\right)
f\left(\fr{y}{\al}\right) L^2(x,y)\, dx dy \nonumber \\
&=& \int_0^1  \int_0^\iy\Biggl(  \int_0^1\biggl(  \int_0^\iy 
x f(\fr{x}{\al}) J_\nu(xt) J_\nu(xt')\,dx\biggr)t' J_\nu(t'y) \,dt'\Biggr)
y f(\fr{y}{\al}) t J_\nu(ty) \, dy dt\nonumber \\
&=& \int_0^1 \int_0^\iy\Biggl(\int_0^\al\biggl(
\int_0^\iy  x f(x) J_\nu(t\al x)
J_\nu(t'x)\, dx\biggr)t' J_\nu(\fr{t'y}{\al})\, dt'\Biggr) y f(\fr{y}{\al}) t
J_\nu(ty)\, dy dt \nonumber \\
\label{I2}
\end{eqnarray}
where we used (\ref{L(x,y)}) to deduce the  middle equality and we made
the change of variables $x/\al\ra x$ and $\al t'\ra t'$ to obtain
the last equality.
\par
We now recall the Hankel inversion formula:
\[ \int_0^\iy u \biggl(\int_0^\iy x g(x) J_\nu(xu)\, dx\biggr) J_\nu(u\xi)\, du
=g(\xi) \] 
which hold for $\sq{x} g(x)$ continuous and absolutely integrable on
the positive real line and $\nu>-\fr{1}{2}$.
First writing the $t'$-integration in (\ref{I2}) as the integral from $(0,\iy)$ 
minus the integral from $(\al,\iy)$ and then  employing the Hankel inversion
formula (with the choice $g(x)=f(x) J_\nu(\al t x)$ )
 on the part containing the $t'$-integration from $(0,\iy)$,  we
see that this part {\it exactly\/} cancels the single integral appearing
in the expression (\ref{var3})  for ${\rm var}~f$.  Thus we are left with 
\begin{eqnarray}
{\rm var}~f &=& \int_0^1 dt\, \int_0^\iy dy\, \int_\al^\iy dt'\,
\int_0^\iy dx\, x y t t' f(x) f(\fr{y}{\al}) J_\nu(\al t x)
J_\nu(\fr{t'y}{\al}) J_\nu(t'x) J_\nu(ty) \nonumber \\
&=& \al^4 \int_0^1 dt\, \int_0^\iy dy\, \int_1^\iy ds\, \int_0^\iy dx\,
x y t s f(x) f(y) J_\nu(\al tx) J_\nu(\al s x) J_\nu(\al s y) J_\nu(\al t y)
\, .
\nonumber \\
&&\label{var4}
\end{eqnarray} 
We remark that the
 Hankel transform  plays the analogous role
 for the Bessel kernel 
that the Fourier transform plays for the sine kernel
\[ \fr{1}{\pi}\fr{\sin\pi(x-y)}{x-y} \]
in the Gaussian Unitary Ensemble.
\subsection{Residue Calculation} 
Introducing  the (inverse) Mellin transform
\[ f(x)=\fr{1}{2\pi i} \int_{c-i\iy}^{c+i\iy} \hat f(z) x^{-z}\, dz
\ \ \ (c>0) \]
into (\ref{var4}) and interchanging the orders of integration we see
that
\begin{eqnarray*}
 {\rm var}~f&=& \al^4 \fr{1}{2\pi i} \int_{c-i\iy}^{c+i\iy} dz_1\,
\hat f(z_1) \fr{1}{2\pi i} \int_{c-i\iy}^{c+i\iy} dz_2\, 
\hat f(z_2) \nonumber \\
&&\times \int_0^1 dt\, t \int_1^\iy ds\, s 
\int_0^\iy dx \, x^{-z_1+1} J_\nu(\al t x) J_\nu(\al s x)
\int_0^\iy dy\, y^{-z_2+1} J_\nu(\al t y) J_\nu(\al s y) \, .
\end{eqnarray*}
The $x$ and $y$ integrations  can be performed using (6.5762) in \cite{GR};
namely,
\begin{eqnarray*}
\int_0^\iy x^{-\la} J_\nu(ax) J_\nu(bx)\, dx
&=&\fr{(ab)^\nu \Gamma\left(\nu+\fr{1-\la}{2}\right)}{2^\la (a+b)^{2\nu-\la+1}
\Gamma\left(1+\nu\right)\Gamma\left(\fr{1+\la}{2}\right)} \\
&&\times F\left(\nu+\fr{1-\la}{2},\nu+\fr{1}{2};2\nu+1;\fr{4ab}{(a+b)^2}\right)
\end{eqnarray*}
where $F(a,b;c;z)$ is the hypergeometric function, $a,b>0$,
$2\Re(\nu)+1>\Re(\la)>-1$. 
\par
In the resulting integral we
 make the following  change of variables:
\[ u=\fr{4ts}{(t+s)^2}\> , \ \ \ v=t+s \]
which has Jacobian
\[  J(u,v)= \fr{v}{4\sq{1-u}}. \]
In the $vu$-plane we are now integrating over the region in the
first quadrant bounded above by the curve
\[ u=\fr{4(v-1)}{v^2}, \ \ \ v\geq 1  \]
and  bounded below by the ray $[1,\iy]$ on the $v$-axis.  The $v$-integration
may now be trivially done with the result that
\begin{eqnarray*}
{\rm var}~f &=& \fr{1}{4^{2\nu-1}\Gamma^2(\nu+1)} \fr{1}{2\pi i}
\int_{c-i\iy}^{c+i\iy} dz_1\, \hat f(z_1)
\al^{z_1}  \fr{\Gamma\left(\nu+1-z_1/2\right)}
{\Gamma\left(z_1/2\right)} \\
&& \times \fr{1}{2\pi i}
\int_{c-i\iy}^{c+i\iy} dz_2\, \hat f(z_2)
\al^{z_2}  \fr{\Gamma\left(\nu+1-z_2/2\right)}
{\Gamma\left(z_2/2\right)} \\
&& \times \fr{1}{z_1+z_2} \int_0^1 u^{2\nu+1-z_1-z_2} \left(1-u\right)^{-1/2}
\left[ \left(1+\sq{1-u}\right)^{z_1+z_2}-
\left(1-\sq{1-u}\right)^{z_1+z_2}\right]\\
&& \times F_\nu(z_1,u) F_\nu(z_2,u)
\end{eqnarray*}
where
\[ F_\nu(z,u):=F\left(\nu+1-\fr{z}{2},\nu+\fr{1}{2};2\nu+1;u\right). \]
We now use $z_1$ and 
 $z=z_1+z_2$ as integration variables so that
\begin{eqnarray*}
{\rm var}~f &=& \fr{1}{4^{2\nu-1}\Gamma^2(\nu+1)} \fr{1}{2\pi i}
\int_{c-i\iy}^{c+i\iy} dz_1\, \hat f(z_1)
  \fr{\Gamma\left(\nu+1-z_1/2\right)}
{\Gamma\left(z_1/2\right)}\\
&& \times \fr{1}{2\pi i}
\int_{2c-i\iy}^{2c+i\iy} dz\, \hat f(z-z_1)
\al^{z}  \fr{\Gamma\left(\nu+1-(z-z_1)/2 \right)}
{\Gamma\left((z-z_1)/2\right)} \\
&& \times  \int_0^1 u^{2\nu+1-z} \left(1-u\right)^{-1/2}
\left[\fr{ \left(1+\sq{1-u}\right)^{z}-
\left(1-\sq{1-u}\right)^{z}}{z}\right]\\
&& \times  F_\nu(z_1,u) \,  F_\nu(z-z_1,u) \, .
\end{eqnarray*}
Observe that the $\al$ dependence of ${\rm var}~f$ resides solely
in the term $\al^z$ in the above integral.  To compute $\al\ra\iy$ this
suggests we should first  deform the contour into the left-half 
$z$-plane.  The $\lim_{\al\ra\iy} {\rm var}~f$ will be determined
by the residue of the pole at $z=0$.
\par
To calculate this residue (which is a function of $z_1$)
 we must know the principal part of the Laurent
expansion (in $z$)  of the integral involving the $u$-integration.  
The divergence of this integral as $z\ra 0$ is determined by the
behavior of the integrand in the vicinity of $u=1$.  This
behavior near $u=1$  is straightforward
to compute since it is known \cite{erdelyi}  that
\[ F(a,b;c,u) \sim \fr{\Gamma(c)\Gamma(a+b-c)}{\Gamma(a)\Gamma(b)}
\left(1-u\right)^{c-a-b} \ \ \ {\rm as} \ \ \ u\ra 1. \]
\par
Thus $\lim_{\al\ra\iy} {\rm var}~f$
 is expressed as a single integral over the variable $z_1$.
If we now make use of the $\Gamma$-function identities \cite{erdelyi}
\begin{eqnarray*}
\Gamma(z)\Gamma(-z)&=& - \fr{\pi}{z\sin\pi z} \, , \\
\Gamma(1/2+z)\Gamma(1/2-z)&=&\fr{\pi}{\cos \pi z}\, , \\
\Gamma(2\nu+1)&=&2^{2\nu}\pi^{-1/2} \Gamma(\nu+1/2)\Gamma(\nu+1) \, ,
\end{eqnarray*}
we obtain
\[ \lim_{\al\ra\iy}{\rm var}~f = -\fr{1}{2\pi^2 i}
\int_{c-i\iy}^{c+i\iy} \hat f(z_1) \hat f(-z_1) \fr{z_1}{2} \tan(\fr{\pi}{2}z_1
) \, dz_1 \, . \]
We now deform the contour  to the imaginary axis (and send $y\ra 2y$) to
obtain (\ref{constant}).

\acknowledgments
The authors wish to thank Professor Harold Widom for his helpful comments.

\end{document}